\documentclass[preprint2]{aastex}

\usepackage{epstopdf}
\usepackage{array}

\shorttitle{X-ray Binary Pulsar OAO 1657-415}
\shortauthors{P. Jenke et al.}

\begin{document}


\title{ORBITAL DECAY AND EVIDENCE OF DISK FORMATION IN THE X-RAY BINARY PULSAR OAO 1657?415}


\author{P. A. Jenke}
\affil{MSFC/NPP, Huntsville, AL 35812, USA}

\author{M. H. Finger}
\affil{Universities Space Research Association, Huntsville, AL 35805, USA}
\author{C. A. Wilson-Hodge}
\affil{Marshall Space Flight Center, Huntsville, AL 35812, USA}

\and

\author{A. Camero-Arranz}
\affil{Universities Space Research Association, Huntsville, AL 35805, USA}




\begin{abstract}
OAO 1657-415 is an eclipsing X-ray binary wind-fed pulsar that has exhibited smooth spin-up/spin-down episodes and has undergone several torque reversals throughout its long history of observation.  We present a frequency history spanning nearly 19 years of observations from the Burst and Transient Source Experiment (CGRO/BATSE) and from the Gamma-Ray Burst Monitor (\emph{Fermi}/GBM).  Our analysis suggests two modes of accretion:  one resulting in steady spin-up correlated with flux during which we believe a stable accretion disk is present and one in which the neutron star is spinning down at a lesser rate which is uncorrelated with flux.  Orbital elements of the pulsar system are determined at several intervals throughout this history.  With these ephemerides, statistically significant orbital decay with a $\dot{P}_{orb} =(-9.74\pm0.78)\times10^{-8}$ is established.

\end{abstract}




\section{Introduction}
OAO 1657-415 is an accreting X-ray pulsar with a high mass companion.  It was first detected by the Copernicus satellite in 1978 \citep{Polidan1978}.  The companion is believed to be a Ofpe/WN9 supergiant \citep{Mason2009}.   OAO 1657-415 is one of 10 known eclipsing X-ray binary pulsars.  Its eclipse lasts 16\% of the orbit or 1.7 days.  OAO 1657-415's moderate spin period (37 s) and its moderate orbital period (10.4 days) \citep{Chakrabarty1993} gives it a unique placement on the Corbet diagram which categorizes accreting pulsars by spin period and orbital period.  Due to the relatively short orbital period and spin period of this wind-fed system  \citet{Mason2009} speculated that this system may be in transition from wind-fed to disk-mediated accretion.   The first part of this work establishes statistically significant orbital decay of the OAO 1657-415 system using data from the Burst and Transient Source Experiment (BATSE) previously on board the Compton Gamma-Ray Observatory (CGRO) and the Gamma-Ray Burst Monitor (GBM) on board \emph{Fermi}. 

 Historically, the supergiant binaries have been subdivided into two groups according to the dominant mode of mass transfer:  Roche lobe overflow or capture from the stellar wind \citep{Bildsten1997}.  This terminology is mismatched in that Roche lobe overflow describes the outflow from the companion while capture from the stellar wind describes the accretion onto the compact object.  An improved classification scheme would be to identify two classes of high mass X-ray pulsar binaries (HMXBs): A) Roche lobe filling supergiants and B) underfilled Roche lobe supergiants.  We present, in the second part of this work, evidence that, for the underfilled Roche lobe systems, accretion can occur either directly from the stellar wind (direct wind accretion) or from an accretion disk created from the stellar wind (disk wind accretion).  The underfilled Roche lobe supergiants may then be further divided into two subclasses: Direct wind accretion and  disk wind accretion.

\section{Background}

\begin{figure}[h!]
   \centering
   \includegraphics[width=2.7in]{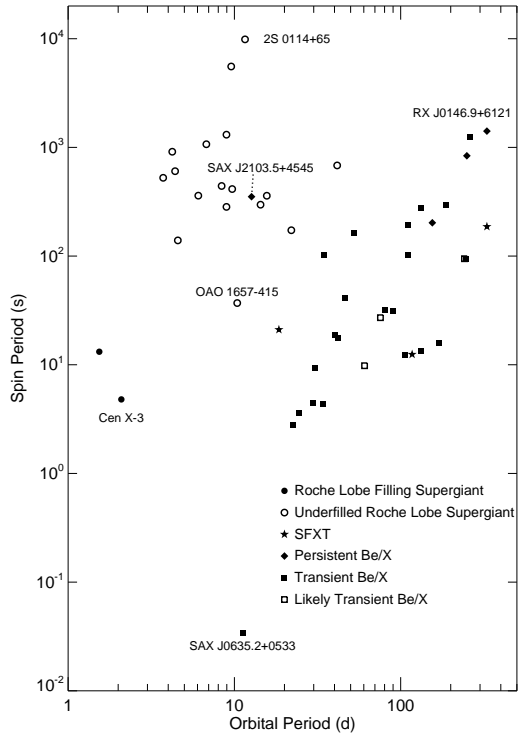}
   \caption{All the Galactic high mass X-ray binary pulsars with known orbital periods plotted on a Corbet diagram.  The hollow circles are the wind-fed supergiants while the filled circles are the Roche-lobe overflow supergiants.  The filled squares are the transient Be/X-ray binaries while the hollow squares are a likely Be X-ray binary.  The filled diamonds are the persistent Be/X-ray star.  The filled stars are the Supergiant Fast X-ray Transient (SFXT).} 
   \label{fig:Corbet}
\end{figure}

	The different classes of HMXBs can be plotted in a P$_{\rm{spin}}$ - P$_{\rm{orb}}$ diagram \citep{Corbet1986}, or a Corbet diagram, where P$_{\rm{spin}}$ is the neutron star spin period and P$_{\rm{orb}}$ is the orbital period of the binary system (see Figure \ref{fig:Corbet}).  

The filled circles are Roche lobe filling high mass systems that have relatively short orbital periods and short spin periods.  The wind fed systems (hollow circles) have longer orbital periods, avoiding Roche lobe filling, and longer spin periods due to weaker accretion torques.   The Be systems (filled squares) display correlation between orbital period and spin period, possibly because the neutron star is further away from the donor star in systems with a long orbital period and it is sampling a weaker portion of the circumstellar disk \citep{Waters1989}.  Other systems show anomalous behavior, such as SAX J2103.5+4545 \citep{Choni2007} which is a Be/X-ray binary with a orbital period of 12.68 days and a pulse period of 358 s.  This unusual orbital and pulse period for a Be/X-ray binary places it among the wind fed supergiants in the Corbet diagram.  OAO 1657-415 also appears to defy these groupings.  With an orbital period of 10.4 days and a pulse period of 37.1 s, it occupies an intermediate region between sources with mass transfer via a stellar wind and Roche lobe overflow \citep{Chakrabarty1993}.   

	Identification of OAO 1657-415's companion proved elusive for 15 years after its  discovery \citep{Kamata1990}.  In 1993, monitoring by CGRO/BATSE allowed sufficient data for the first orbital ephemeris to be calculated yielding a mass function for the system  \citep{Chakrabarty1993}.  Assuming the pulsar has a mass of 1.4 M$_{\sun}$ and using the knowledge of the eclipse, \citet{Chakrabarty1993} calculated the mass of the companion to be 14 M$_{\sun}$ $\leq$ M$_{c}$ $\leq$ 18 M$_{\sun}$ with a radius between 25 and 32 R$_{\sun}$.  This  yielded a blue supergiant of stellar class B0-B6.  Attempts to optically identify the companion failed due to a large column density but in 2002 a relatively bright star (2MASSJ17004888-4139214) was identified in the near-infrared that was coincident with a \emph{Chandra X-Ray Observatory} position of OAO 1657-415  \citep{Chakrabarty2002} and was consistent with its companion's spectral class.  Due to the degeneracy of infrared color-color diagrams, infrared photometry did not provide decisive results.  More detailed infrared spectroscopy performed in 2009 by \citet{Mason2009} concluded that OAO 1657-415' s companion is of a spectral type Ofpe/WN9, which is a transitional stage between a type OB and a Wolf-Rayet star characterized by high mass loss, low wind velocities, and exposed CNO-cycle products.
\section{Observations}
BATSE was an all-sky monitor designed to study gamma-ray bursts and transient source outbursts in the hard X-ray and soft gamma-ray bands \citep{Fishman1989}. The observations reported here use data from the eight Large Area Detectors (LADs), which were 2025 cm$^{2}$ in area by 1.24 cm thick Na I(Tl) scintillators operated in the 20 keV-1.8 MeV band. These were located at the eight corners of the Compton Gamma Ray Observatory (CGRO). The LADs were uncollimated, with each detector viewing half of the sky. The pulse timing analyses uses the DISCLA channel 1 data, which consist of discriminator rates for the 20-50 keV band, continuously read out from all eight detectors with a resolution of 1.024 s.  BATSE data used for the following analysis spans from  MJD 48362 to 50683 (1991 April 16 - 1997 August 23).

	 GBM is an all sky monitor whose primary objective is to extend the energy range over which gamma-ray bursts are observed in the Large Area Telescope (LAT) on \emph{Fermi} \citep{Meegan2009}.  GBM consists of 12 NaI detectors with a diameter of 12.7 cm and a thickness of 1.27 cm and two BGO detectors with a diameter and thickness of 12.7 cm.  The NaI detectors have an energy range from 8 keV to 1 MeV while the BGO's extend the energy range to 40 MeV.  The current pulse timing analysis uses the first 3 channels of the CTIME data (8-12 keV, 12-25 keV, 25-50 keV) from the NaI detectors with a time resolution of 0.256 s.  The daily light curves are visually inspected and times where the spacecraft is performing a rapid maneuver  or passing over the South Atlantic Anomaly are removed.  Other contaminating features that prevent a good background fit, such as gamma-ray bursts or solar flares, are also removed.  The data from channel 0-2 (corresponding to energies between 8 and 50 keV) are fit with an empirical background model which is subtracted from the data.  GBM data used for the following analysis spans from  MJD 54690 to 55824 (2008 August 12 - 2011 September 20).

 \emph{Swift}/BAT is a hard X-ray monitor that has a field-of-view of 1.4 steradians and its  array of CdZnTe detectors are sensitive in 15 - 150 keV range.  It is a coded aperture instrument with a detector area of 5200 cm$^{2}$.  We use the \emph{Swift}/BAT transient monitor results\footnote{\url{http://heasarc.gsfc.nasa.gov/docs/swift/results/transients/}}  (15-50 keV), provided by the \emph{Swift}/BAT team, in order to monitor the  total flux for OAO 1657-415 from  MJD 54690 to 55824 (2008 August 12 - 2011 September 20).  
 

\section{Data Analysis}
 Because pulsars emit a large percentage of their radiation in periodic pulses, Fourier components of the source signal can be identified and extracted from the background.  For details on this procedure see \cite{Camero2010,Finger1999}.   In the case of BATSE data, short intervals (roughly 500 s) were fit to a Fourier expansion in pulse phase plus a quadratic spline model that accounts for the combined background and the average source flux.  In the case of GBM data, short intervals (again roughly 500 s) of the residuals from the background subtraction were fit to a Fourier expansion in pulse phase.  OAO 1657-415 pulse profiles, from BATSE and GBM data,  were represented by six harmonics $(n=6)$ in this work.  Additional harmonics did not result in statistically significant improvements in pulse structure. The Fourier expansion in pulse phase resulted in an estimated pulse profile (residual profile) for each data segment.  For the purpose of creating a preliminary phase model, the occulted intervals were removed and the remainder of the  orbit was initially divided into threeeequally spaced intervals. 
 
	The times are barycentered using the JPL Planetary ephemeris DE200 \citep{Standish1990}.  Further time corrections are performed  to remove the time delays from the orbital motion of the pulsar using a preliminary orbital model.  This results in times that correspond to the pulsar emission time  $t^{em}$  that are computed from barycentered times ($t^{bc}$) by $t^{em}  = t^{bc}-z$. The line of sight delay, $z$, associated with the binary orbit of the pulsar is given by \citet{Deeter1981} 
\begin{eqnarray}
\nonumber z(t^{em}) = & a_x \sin{i} [\sin{\omega} (\cos{E} - e) + (1 - e^2)^{\frac{1}{2}}  \\
                                        & \times\cos{\omega} \sin{E}] 
\label{equ:z}
\end{eqnarray}
where $a_{x}$ is the projected semi-major axis of the pulsar's orbit, $i$ is the orbit's inclination
relative to the plane of the sky, $\omega$ is the periastron angle, and $e$ is the orbital eccentricity. The
eccentric anomaly, $E$, is related to time through Kepler's equation,
\begin{equation}
E - e \sin{E} = (\frac{2 \pi} { P_{orb}} )(t^{em} - \tau_p) 
\end{equation}
where $P_{orb}$ is the orbital period and $\tau_p$ is the periastron epoch, the time at which the pulsar most closely approaches its companion.  Initially, the orbital ephemeris from \citet{Bildsten1997} is used to correct the times for orbital motion.  

\begin{figure}[h]
\centering
  \includegraphics[width=3in]{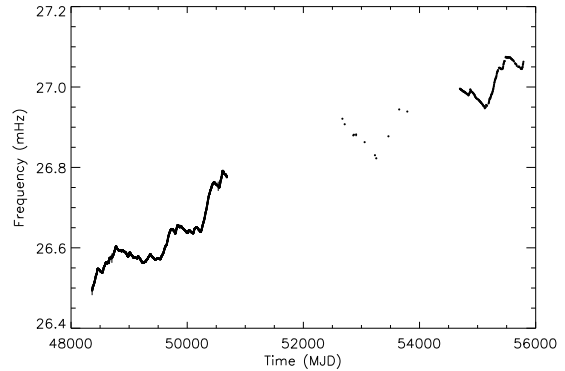}
  \caption{Frequency history of OAO 1657-415 extracted from BATSE data (left), \emph{INTEGRAL} data from  \citet{Barnstedt2008} (center), and GBM data (right).  }  
  \label{frequency}
\end{figure}

	A preliminary phase model was estimated from historical trends in the pulse frequency.  The pulse phase model is a monotonically increasing function counting the number of pulses from the pulsar since a defined time $\tau$ referred to as the epoch.  A simple phase model is:
\begin{equation}
		\phi(t_k) = \phi_o + \nu_{o}(t_k-\tau) +\dot{\nu}_{o} \frac{(t_k-\tau)^2}{2} 
\end{equation}

where $\nu_{o}$ is the pulse frequency at time $\tau$ and $\phi_{o}$ is the phase at time $\tau$. The pulse frequency, $\nu(t_{k})$ is the first derivative of the phase model.  In the simple case, $\nu(t_{k})= \nu_{o}$, where $\nu_{o}$ is a constant and $\dot{\nu}_{o} = 0$.  As more knowledge was gained about the pulsar's behavior, the phase model was refined by adding additional terms.  The long term phase model was determined by fitting quadratic splines to the corrected phases.  Each unocculted interval was searched for a pulsed signal, in a narrow frequency band, using the  $Y_{n}$ statistic described in \citet{Finger1999}.  The search was performed on the first two harmonics and using channels 1+2 (12-50 keV) simultaneously for GBM and channel 1 (20-50 keV) for BATSE data.  The preliminary frequency history of OAO 1657-415 is shown in Figure \ref{frequency} with a long term spin-up trend ($\dot{\nu} \sim 7.15 \times 10^{-8}$ Hz day$^{-1}$) being evident.  Steady spin-up is evidence for a prograde accretion disk.

\begin{figure}[h]
\centering
  \includegraphics[width=3in]{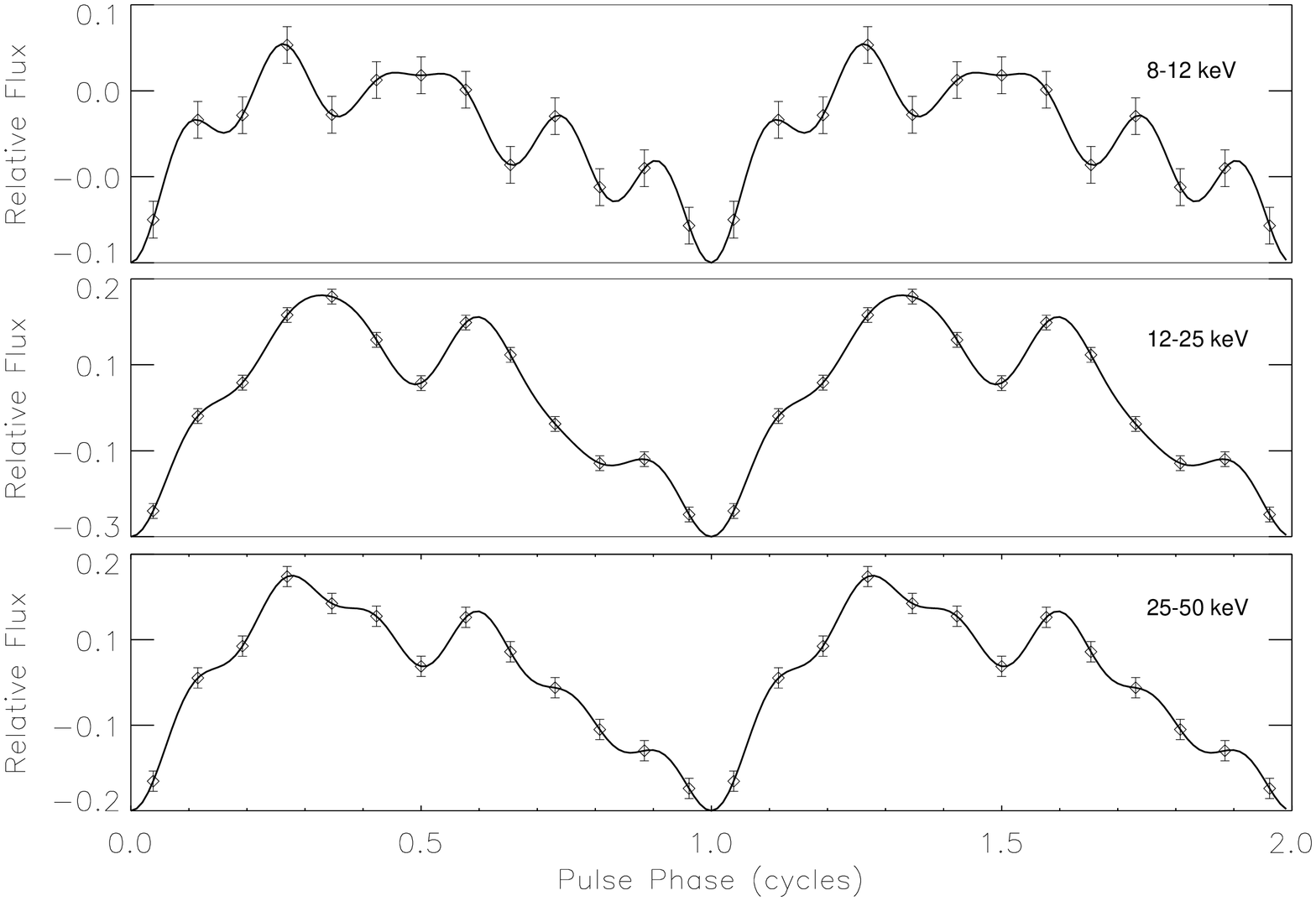}
  \caption{Average pulse profiles extracted from GBM data for 10orbits of  OAO 1657-415  for 8-12 keV (top panel), 12-25 keV (middle panel), and 25-50 keV (bottom panel).  Two cycles are shown for clarity.   The points with error bars are approximately statistically independent} \label{pulsevsenergy}
\end{figure}

\begin{figure}[h]
\centering
  \includegraphics[width=3in]{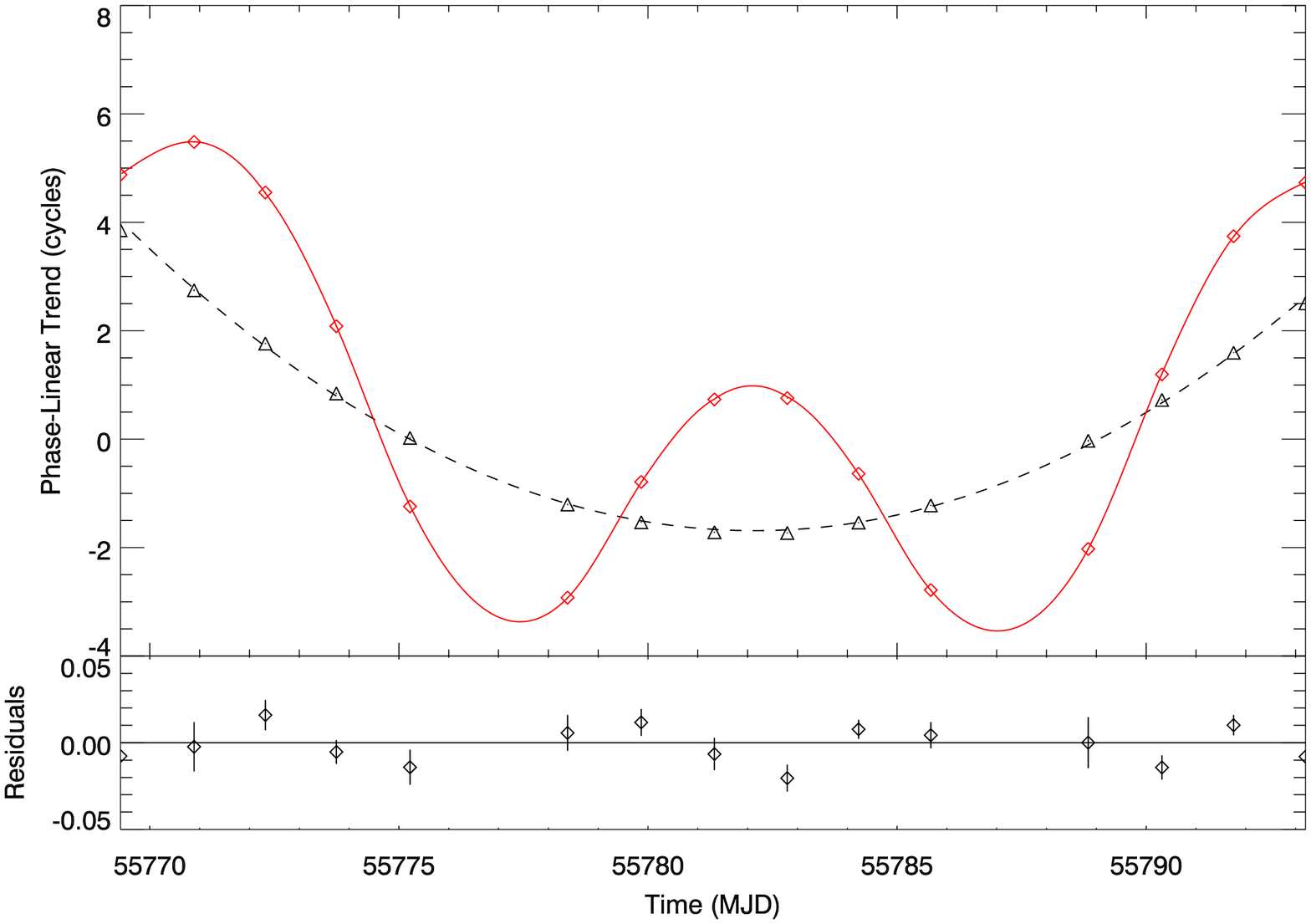}
  \caption{Black \bf{dashed} curve is the pulse phase model minus a linear trend in emission time for one three orbit fit.  The black triangles are the measured phases using times corrected for the estimated orbit with the same linear trend in barycentered time removed.  The red diamonds are the same phases with the same linear trend removed using times not orbitally corrected.  The red curve is the best fit model of the phases  with the same linear trend removed.  The lower panel displays the residual to the fit of the orbital model.}
\label{model}
\end{figure}

	In order to determine the most accurate phases,  the occulted intervals were removed and the remainder of the orbits were divided into six intervals.  The residual profiles were folded for each of these intervals using the most recent ephemeris from \citet{Bildsten1997} and the best frequency from the above search.  Although no frequency search was performed at this stage, the  $Y_{2}$ statistic was calculated for the previously determined frequency for GBM data.  Templates for the pulse profile were calculated by averaging the folded residual profiles over three orbit intervals.  For GBM data, the 12-25 keV energy band was used.  Long term average pulse profiles from GBM data in three energy bands (8-12 keV, 12-25 keV, and 25-50 keV) are shown in Figure \ref{pulsevsenergy}.  The morphology of the pulse changes from a relatively smooth single pulse at low energy to a double peaked asymmetric pulse with a fast rise and slow decay at higher energies.  There was very little evolution of the pulse profile with luminosity or time.  The 12-25 keV pulse profile is representative of the template profiles used.  The template profiles were normalized and fit to the folded profiles along with a phase model.  A scaling factor and phase offset was calculated from the fit.  For BATSE data, phase offsets that could unambiguously determine the phase to ~3$\sigma$ were selected (i.e. phase offsets  with error $\le$ 0.15).   For GBM data, a cut in $Y_{2}$ ($\ge$ 8)statistic was used to select good intervals.  The phase model was then corrected for the fitted phase offsets to produce the final corrected phases.  
	The corrected phases and barycentered times were fit to an orbital model plus a polynomial (see Figure \ref{model}).  The first term of the polynomial is the linear trend in time expected from the phases while the second term represents the frequency change of the source over the interval.  Higher order terms account for torque noise inherent in these systems.
%

\section{Timing Results}

\begin{figure}[h]
\centering
  \includegraphics[width=3in]{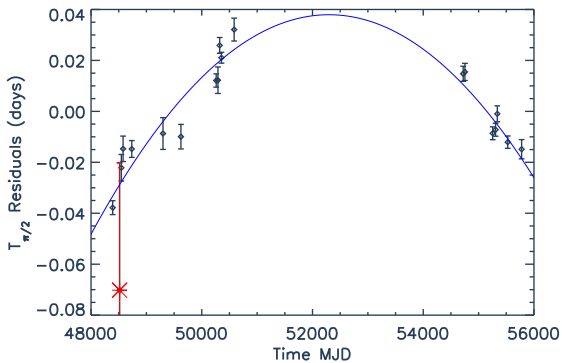}
  \caption{Diamonds are orbital epochs for each ephemeris with a linear trend removed.  The line is a quadratic fit to the epochs.  The epochs to the left are calculated from BATSE data while the epochs on the right are calculated from GBM data.  The epoch calculated by \citet{Chakrabarty1993} is labeled as a red star and is excluded in the quadratic fit.}  \label{decay}
\end{figure}

Intervals where pulsations were clearly detected and where the phases did not deviate radically from the phase model (minimal torque noise) were chosen from BATSE and GBM data.  Orbital fits to  these data were performed for three orbit intervals (the same intervals in which the pulse profile templates were developed) from the selected data.  Table \ref{tbl-1} shows the intervals and results of each fit.  A new orbital ephemeris was determined by averaging the orbital elements (with the exception of the orbital period, orbital period derivative and epoch) from each fit  (See Table \ref{tbl-2}).
\begin{table*}[h!]
\begin{center}
\setlength{\extrarowheight}{2pt}
\caption{Orbital Ephemerides for OAO 1657-415.\label{tbl-1}}
\begin{tabular}{lcccccccc}
\tableline\tableline
Start MJD & $T_{\pi/2}$\tablenotemark{a} [MJD]  & $a_{x}\sin{i}$\tablenotemark{a}  [lt-sec] & $e\cos\omega$  & $e\sin\omega$  &  $\chi^{2}$ & DOF \\
\tableline
48370.7  &48390.6549$\pm$0.0027  &106.49$\pm$0.29  &   -0.0053$\pm$0.0026  & 0.1151$\pm$0.0038  & 9.8  & 8 \\
48527.5  &48547.3800$\pm$0.0052  &104.01$\pm$0.61  &   -0.0071$\pm$0.0027  & 0.1120$\pm$0.0046  &23.9  & 8 \\
48558.8  &48578.7293$\pm$0.0050  &106.05$\pm$0.47  &   -0.0011$\pm$0.0032  & 0.1047$\pm$0.0050  &14.8  & 8 \\
48715.5  &48735.4386$\pm$0.0033  &106.34$\pm$0.35  &   -0.0032$\pm$0.0035  & 0.1124$\pm$0.0057  & 7.6  & 8 \\
49279.7  &49299.5984$\pm$0.0062  &105.09$\pm$0.60  &   -0.0069$\pm$0.0037  & 0.0999$\pm$0.0052  &23.8  & 8 \\
49603.6  &49623.4633$\pm$0.0048  &106.42$\pm$0.52  &   -0.0096$\pm$0.0034  & 0.1101$\pm$0.0052  &18.7  & 8 \\
50240.9  &50260.7701$\pm$0.0026  &106.16$\pm$0.29  &   -0.0053$\pm$0.0022  & 0.1061$\pm$0.0032  &12.3  & 8 \\
50272.2  &50292.1121$\pm$0.0052  &106.82$\pm$0.56  &   -0.0098$\pm$0.0036  & 0.1140$\pm$0.0056  &21.5  & 8 \\
50303.5  &50323.4677$\pm$0.0031  &106.90$\pm$0.34  &    0.0061$\pm$0.0032  & 0.1071$\pm$0.0048  & 1.5  & 8 \\
50334.9  &50354.8047$\pm$0.0021  &105.40$\pm$0.23  &   -0.0042$\pm$0.0021  & 0.1047$\pm$0.0032  & 6.8  & 8 \\
50564.7  &50584.6562$\pm$0.0045  &107.08$\pm$0.51  &   -0.0035$\pm$0.0031  & 0.1162$\pm$0.0050  &16.4  & 8 \\
54700.0  &54721.7666$\pm$0.0030  &106.30$\pm$0.30  &   -0.0033$\pm$0.0027  & 0.1045$\pm$0.0048  & 3.4  & 8 \\
54732.0  &54753.1092$\pm$0.0034  &105.74$\pm$0.36  &   -0.0086$\pm$0.0031  & 0.1018$\pm$0.0044  &12.2  & 9 \\
55243.0  &55254.5552$\pm$0.0025  &105.84$\pm$0.30  &   -0.0113$\pm$0.0024  & 0.1015$\pm$0.0036  &14.4  & 9 \\
55286.0  &55306.7930$\pm$0.0025  &106.31$\pm$0.28  &   -0.0056$\pm$0.0022  & 0.1021$\pm$0.0036  &14.8  & 9 \\
55317.0  &55338.1411$\pm$0.0031  &107.00$\pm$0.44  &   -0.0055$\pm$0.0024  & 0.1114$\pm$0.0038  &23.0  & 9 \\
55505.0  &55526.1813$\pm$0.0024  &106.36$\pm$0.28  &    0.0010$\pm$0.0025  & 0.1098$\pm$0.0040  & 8.4  & 9 \\
55765.0  &55776.9135$\pm$0.0038  &106.52$\pm$0.53  &   -0.0064$\pm$0.0024  & 0.1071$\pm$0.0036  &26.5  & 6 \\
\tableline
\end{tabular}
\tablenotetext{a}{Errors are statistical and inflated to reflect $\chi^{2}_{\nu}$.}
\end{center}
\end{table*}
\begin{table}[h!]
\begin{center}
\caption{\small{Average Orbital Elements for OAO~1657-415.\label{tbl-2}}}
\scalebox{0.77}{
\begin{tabular}{lcc}
\tableline\tableline
Epoch &$T_{o}$ & 52298.01(79) MJD\\
Orbital Period  & $P_{orb}$ &  10.44729(21) days\\
Orbital Decay  & $\dot{P}_{orb}$ &$(-9.74 \pm0.78)\times10^{-8}$ \\
Proj. Semimajor axis & $a_{x}\sin{i}$ &        106.157$\pm$    0.083 lt-sec\\
Long. of Periastron & $\omega$ &        92.69$^{\circ} \pm$      0.67$^{\circ} $\\
Eccentricity & $e$ &       0.1075$\pm$   0.0012\\
Pulsar Mass Function & $f_{x}(M)$ &        11.473$\pm$     0.027 M$_{\odot}$\\
\tableline
\end{tabular}
}
\label{table}
\end{center}
\end{table}
All the orbital epochs are plotted  against time with a linear trend removed (see Figure \ref{decay}).
  A quadratic fit in orbit number (see Figure \ref{decay})  from all the determined epochs reveals the long term orbital trend in the orbit of OAO 1657-415.  The \citet{Chakrabarty1993} epoch is excluded in this calculation because the data they used overlaps the data used here.  Nevertheless, the \citet{Chakrabarty1993} epoch is consistent with the these results.  A second order expansion of the epoch in orbit number, 
\begin{eqnarray}
T_{\pi/2} = T_{o} + nP_{orb} + \frac{1}{2}n^{2}P_{orb}\dot{P}_{orb},
\end{eqnarray}
along with the quadratic fit above determines the orbital decay, orbital period, and epoch for this system (See Table \ref{table}).
Equating the quadratic term in the expansion of the epoch with the quadratic term in the fit to the epoch (Figure  \ref{decay}) gives  a $\dot{P}_{orb} = (-9.74 \pm 0.78) \times 10^{-8}$.  The inclusion of the epoch calculated from \emph{INTEGRAL} data by \citet{Barnstedt2008} was rejected on the basis that it significantly reduces the quality of the quadratic fit ($\Delta\chi^{2} > 20,000$) in Figure \ref{decay}.  This suggests that the error in the \citet{Barnstedt2008} epoch is significantly underestimated.

\section{Torque-Flux Correlation Results}
\begin{figure}[h]
\centering
  \includegraphics[width=3in]{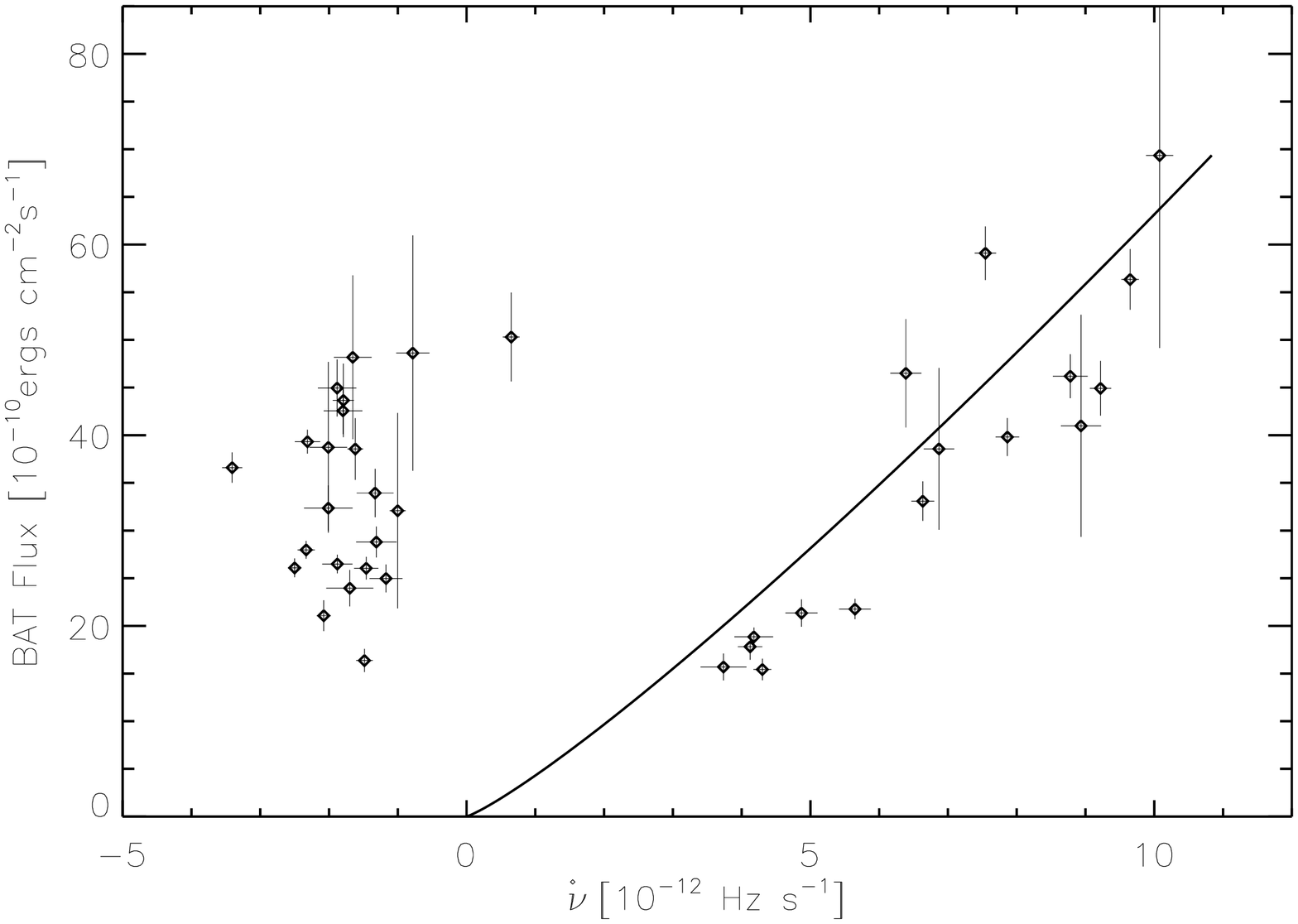}
  \caption{ \emph{Swift}/BAT flux vs. the spin-up of OAO 1657-415 measured using GBM data.  The curve is a model ($\dot{\nu} \propto F^{\frac{6}{7}}$) that is fit to the data where spin-up  is greater than $3\times 10^{-12}$ Hz s$^{-1}$.}  
  \label{BAT_fdot}
\end{figure}

We compare our measured frequency rate ($\dot{\nu}$) with \emph{Swift}/BAT fluxes (Figure \ref{BAT_fdot}).  In order to determine a calibration for the BAT rates, observations ($\sim36$ ks) of OAO 1657-415 by {\it RXTE} were obtained on  2011 August 11.  The {\it RXTE}/PCU2 (5--50 keV)  spectra were fitted in XSPEC 12.7\footnote{Goddard Space Flight Center.   Arnaud K.A, Dorman  B, Gordon C, HEASARC Software Development}  with a model  that includes a low-energy absorption, a power law, the local model fdcut \citep{Tanaka1986}, an iron line and and a few residual lines, in XSPEC notation phabs$\times$($\sum$gaussian+powerlaw)$\times$fdcut.  The energy flux from 5-50 keV during these observations is $4.15\times10^{-9}$erg cm$^{-2}$s$^{-1}$.  Comparing this energy flux and the rates from the \emph{Swift}/BAT transient monitor (15-50 keV) for approximately the same times, a conversion factor of $9.5133\times10^{-8}$ ergs counts$^{-1}$ was determined.  Rates from observations of OAO 1657-415 by the \emph{Swift}/BAT transient monitor were folded on OAO 1657-415's orbital period and converted to fluxes using the above conversion factor.  The maximum flux during an orbital cycle was chosen as a proxy for the intrinsic flux of the source.  This is because the orbital flux profiles suggests OAO 1657-415 is partially obscured during much of its orbit.  This is consistent with locally large column densities inferred for the source \citep{Naik2009,Audley2006}.  The maximum of the folded fluxes during the intervals in which the derivative of the frequency was determined is plotted against the frequency derivative and is shown in Figure  \ref{BAT_fdot}.  The intervals where the frequencies were not monotonically increasing or decreasing were rejected to insure accurate determination of the frequency derivative.   The intervals were at least 12 days to ensure the maximum flux during an orbital period could be determined.  Figure \ref{BAT_fdot} shows a strong flux-$\dot{\nu}$ correlation when $\dot{\nu}$ is greater than $3\times10^{-12}$ Hz s$^{-1}$.  
%

\section{Orbital Decay Discussion}
Orbital decay provides evidence that the system is evolving toward Roche lobe over-flow.  Using values for the Roche lobe radius ($R_{L}$) and the masses of the neutron star and companion from \citet{Mason2012},  we predict that OAO 1657-415 companion will fill its Roche lobe in (8-17)$\times 10^{4}$ years.  This is contingent on the companion maintaining its current evolutionary pattern for that length of time.  If OAO 1657-415's companion is experiencing the onset of a Wolf-Rayet stage then it is probable that it will maintain its current mass loss rate and OAO 1657-415 will evolve into a Roche lobe filling system with a Wolf-Rayet star given that the Wolf-Rayet phase lasts (3-5)$\times 10^{5}$ years \citep{Prantzos1986}.

	A strong stellar wind will transfer angular momentum from the system to a halo surrounding the binary system which will lead to orbital decay.  Following \citet{Heuvel}, the system may be viewed as a massive star loosing material, via stellar wind, to a ring beyond the $L_{2}$ point of the system.  If $a$ is the semimajor axis of the orbit then $a_{\epsilon}$ is the radius of the ring of escaping material.  In this case $a_{\epsilon}$  must be larger than the L$_{2}$ point which is roughly located at $1.2a$ thus $a_{\epsilon}> 1.2 a$.  The ratio of escaping angular momentum per unit mass to total angular momentum per unit mass is
\begin{eqnarray}
\gamma = \frac{(M_c + M_p)^2}{M_c M_p}\left(\frac{a_{\epsilon}}{a(1-e^2)} \right)^{\frac{1}{2}},
\end{eqnarray}
where $M_{c}$ is the mass of the companion, $M_{p}$ is the mass of the neutron star, $e$ is the eccentricity of the orbit. The predicted orbital decay from mass loss through a stellar wind is 
\begin{eqnarray}
\nonumber -\frac{\dot{P}_{orb}}{P_{orb}} = &-(1+3\gamma)\frac{\dot{M}_c +\dot{M}_p}{M_c + M_p} - 3\left(\frac{\dot{M}_c}{M_c}+\frac{\dot{M}_p}{M_p}\right)\\
                                                       &+\frac{3e\dot{e}}{(1-e^2)}.
\end{eqnarray}Ä
Assuming the eccentricity changes slowly and  $|\dot{M}_{c}| \gg |\dot{M}_{p}|$ and $|\dot{M}_{c}/M_{c}| \gg |\dot{M}_{p}/M_{p}|$ then
\begin{eqnarray}
-\frac{\dot{P}_{orb}}{P_{orb}} = -(1+3\gamma)\frac{\dot{M}_c}{M_c + M_p} - 3\frac{\dot{M}_c}{M_c}.
\end{eqnarray}
 Using $M_{p} \le 1.8 $ M$_{\odot}$, and assuming $M_{c} \le$ 18 M$_{\odot}$, and $|\dot{M}_{c}| > 1.0\times10^{-6}$ M$_{\odot}$yr$^{-1}$  from \citet{Mason2012} then
\begin{eqnarray}
-\frac{\dot{P}_{orb}}{P_{orb}}  > 2.0\times10^{-6} \ \mathrm{yr}^{-1},
\end{eqnarray}    
where $P_{orb}$ is the period of the orbit. This results in   -$\dot{P}_{orb} > 5.9 \times10^{-8}$.   This calculation places a lower limit on the angular momentum transferred from the companion to the halo via a stellar wind.  Although spin-orbit quadrupole coupling \citep{Lai1995} almost assuredly plays a role in orbital decay,   the measured orbital decay for OAO 1657-415 is consistent with the predicted orbital decay from stellar wind mass loss.

%
%
%
\section{Flux-Torque Correlation Discussion}

	Flux-torque correlations are expected when an accretion disk is present.  To test for the existence of an accretion disk assuming a high $\dot{M}$,   a model of disk accretion $\dot{\nu} \propto F^{\frac{6}{7}}$ \citep{Rappaport1977} is fit to the data where $\dot{\nu} > 3\times10^{-12}$ Hz s$^{-1}$ (see Figure \ref{BAT_fdot}).  Using this model, the resultant fit, and a range of values for the other parameters, a distance to the OAO 1657-415 can be calculated.   From \citet{Ghosh&Lamb3}, the neutron star spin  frequency derivative, $\dot{\nu}$, may be calculated in terms of the moment of inertia, $I$, mass of the neutron star, $M_{p}$, mass accretion rate, $\dot{M}_{p}$, the radius of the inner point of the  Keplerian disk, $r_o$, and a function,  $n(\omega_{s})$, which is of order unity and only depends on the fastness parameter, $\omega_{s=0}$:

\begin{eqnarray}
\dot{\nu} = n(\omega_o)(2 \pi I)^{-1}(G M_p r_o)^{1/2} \dot{M}_p,
\end{eqnarray}          
were $G$ is the gravitational constant.  Following \citet{Wang1996} the inner point of the Keplerian disk is given by:

\begin{eqnarray}
r_o=k(2 G M_{p})^{-1/7}\mu^{4/7} {\dot{M}_p}^{-2/7},
\end{eqnarray}          
where $\mu$ is the magnetic dipole moment and $k$ is such that $r_{o} = kr_{a}$ where $r_a$ is the Alfven radius for spherical accretion. 
 Assuming at least partial threading of the accretion disk by the magnetosphere, $k$ is less than unity.  
 
\begin{figure}[h]
\centering
  \includegraphics[width=3in]{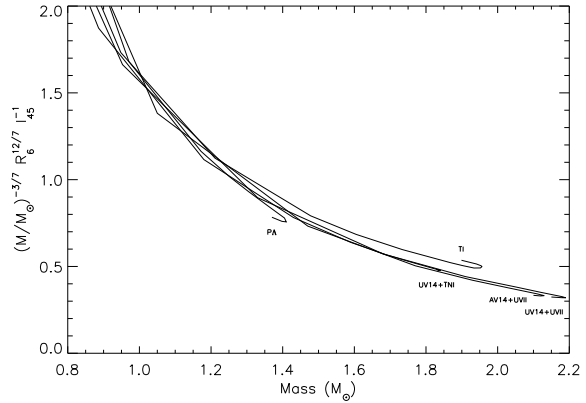}
  \caption{ Models of the mass of a neutron star verses  the combination of parameters, $(M/M_{\odot})^{-3/7}{R_{6}}^{12/7}{I_{45}}^{-1}$ \citep{Wiringa,Arnett,Pandharipande}.  The type of model is insensitive to this grouping of parameters.}
  \label{ns_models}
\end{figure}

 	Assuming a dipole magnetic field and that the gravitational energy of the accreted material is converted to X-rays i.e. $F = \alpha\frac{GM_{p}}{R_{p}}\frac{\dot{M_{p}}}{4 \pi d^2}$ then,
\begin{eqnarray}
\nonumber & \dot{\nu} = 2.46\times10^{-6}(n(\omega_o)k)^{1/2}\alpha^{-6/7}\left(\frac{M_p}{M_\odot}\right)^{-3/7} \\
\nonumber                      &\times \left(\frac{I}{10^{45} \mathrm{ \ g\ cm}^{2}}\right)^{-1} \left(\frac{R_p}{10^6 \mathrm{\ cm}}\right)^{12/7}\\
&\times \left(\frac{B}{10^{12} \mathrm{ \ gauss}}\right)^{2/7} \left(\frac{d}{\mathrm{cm}}\right)^{12/7} \left(\frac{F}{\mathrm{ ergs \ cm}^2\mathrm{\ s}^{-1}}\right)^{6/7}
\label{fdot}
\end{eqnarray}         
 
where $\alpha$ is the bolometric fraction, $R_{p}$ is the neutron star radius, $d$ is the distance to the neutron star in cm, and $B$ is the magnetic field of the neutron star.  By choosing a modest range of possible neutron star masses, $1 \le M_{p} \le 1.8$, models of the equation of state of neutron stars determine $M_{p}$, $R_{p}$, and $I$.  From Figure \ref{ns_models}, this range of neutron star masses results in $0.5 \le (M/M_{\odot})^{-3/7}{R_{6}}^{12/7}{I_{45}}^{-1} \le 1.6$ \citep{Wiringa,Arnett,Pandharipande}  where $R_{6}$ is in units of $10^{6}$ cm and $I_{45}$ is in units of $10^{45}$ g cm$^{2}$.  The model of the energy spectrum was used to infer the bolometric fraction of the luminosity from 5 - 50 keV.  For OAO 1657-415 it was determined that  $0.5 \le \alpha \le 0.8$.  Although a cyclotron resonance scattering feature (cyclotron line) has never been detected for OAO 1657-415, the energy of the cyclotron lines in other X-ray binaries range between 10 - 60 keV.  Broadening this range to 10 - 100 keV implies a magnetic field strength range for this population to be $1.0 \le B_{12} \le 10.0$ G.  The factor, $(n(\omega_o)k)^{1/2}$, is assumed for this work, to be between 0.8 and 1.0.   Using the fitted results in Figure \ref{BAT_fdot} and Equation \ref{fdot}, the distance to OAO 1657-415 can be constrained to between 3 and 16 kpc which is consistent with \citet{Mason2009}, $4.4 \le d \le 12$ kpc.  The existence of a long lived (weeks to months) transient  accretion disk is consistent with the flux-$\dot{\nu}$ correlation and the distance to the star.

	A question of considerable importance to these type of systems is how a stellar wind can transfer nearly uniform (averaged over a few days) angular momentum to the neutron star for periods of months.  Even considering a stable wind emanating from the companion which is unlikely \citep{Naik2009},  the disrupting presence of a neutron star and its accompanying X-rays irradiating the wind and the companion will add considerable complexity to the system.  Aside from the short term periods of spin-up and spin-down, the long term decrease in spin period that has occurred since the discovery of the source suggests that the torque on the neutron star does not stem from stochastic processes.  
	
	 Although our temporal resolution for torque switching is a few days, Figure \ref{BAT_fdot} is reminiscent of an unstable prograde and retrograde transient (hours) accretion disks, described by \citet{Taam&Fryxell1989}, forming  around OAO 1657-415 contributing to the torque responsible for $\dot{\nu} < 3 \times10^{-12}$ Hz s$^{-1}$.  We have concluded that an accretion disk is present during the periods in which $\dot{\nu} > 3 \times 10^{-12}$ hz/s.  These periods last for weeks to months.  It is possible that the disk is forming and disappearing on shorter timescales than this but our technique prevents us from determining torque-luminosity correlations on timescales shorter than the orbital period. 
	
\begin{figure}[h]
\centering
  \includegraphics[width=3in]{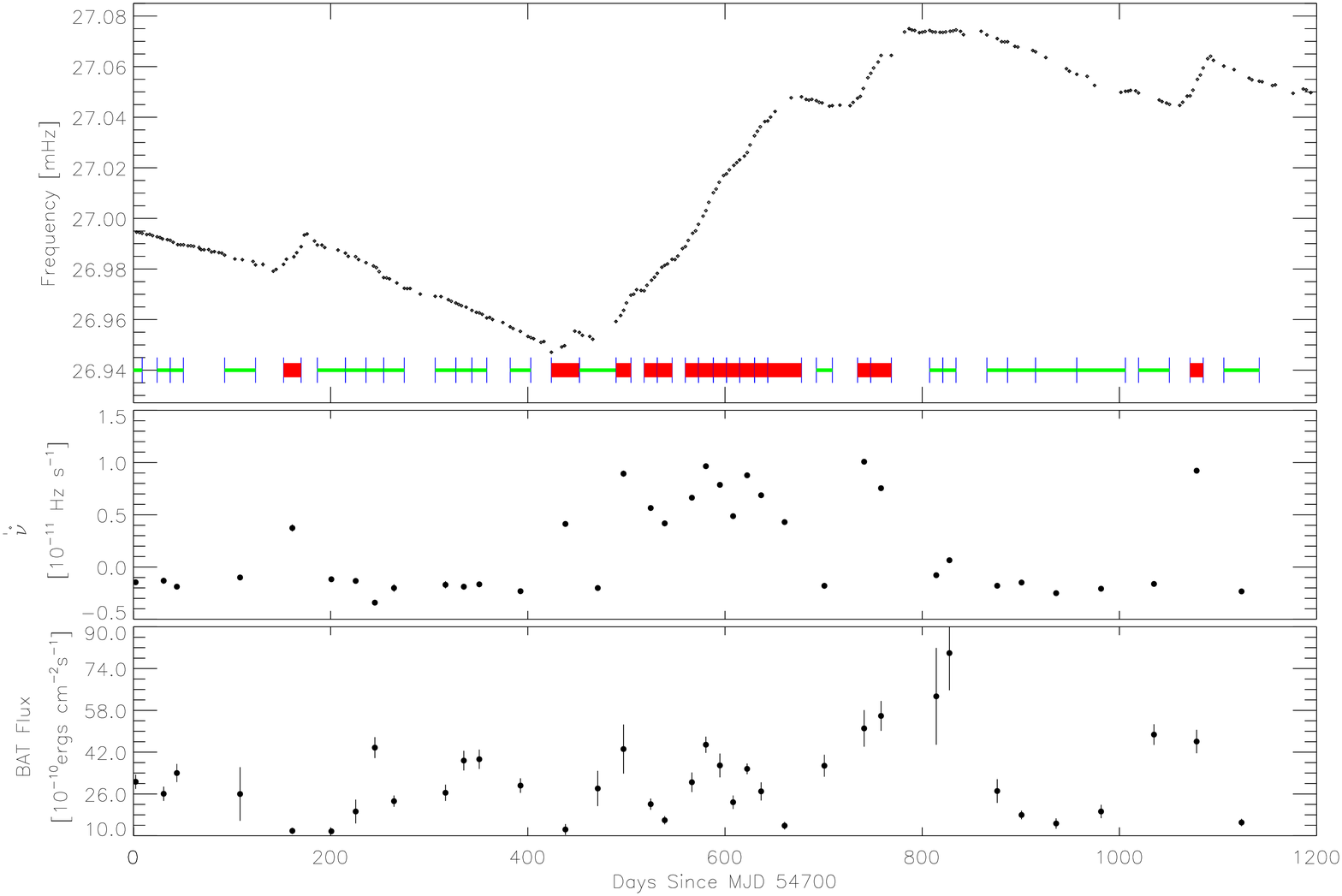}
  \caption{Top panel is a frequency history of OAO 1657-415 showing, at the bottom of the panel, the intervals where the frequency derivative was calculated.  The intervals are travesed by a narrow light gray line at times where the frequency derivative and flux are not correlated.  They are traversed by a wide dark gray line where the frequency derivative and flux are correlated.  The middle panel shows the calculated frequency derivative vs. time for the above intervals while the bottom panel shows the maximum of the \emph{Swift}/BAT orbitally folded flux [$10^{-10}$ erg cm$^{-2}$ s$^{-1}$] for the same intervals. }
  \label{BAT_fdot_timing}
\end{figure}

	Figure \ref{BAT_fdot_timing} shows the frequency, frequency derivative, and the maximum of the orbital folded flux vs. time.  During spin-up, OAO 1657-415 is more likely to exhibit torque-luminosity correlations which are not apparent while the source is spinning down.
 	The correlated part of Figure \ref{BAT_fdot} ($\dot{\nu} > 3 \times 10^{-12}$ hz/s) spans the same flux range as the uncorrelated part, therefore the flux alone is not a good proxy for the torque applied to the neutron star (See the third panel in figure \ref{BAT_fdot_timing}).  If this is the case then the formation of the stable prograde accretion disk is not triggered by high $\dot{M}$ but must have some other, unknown, triggering mechanism.  Additional three dimensional hydrodynamic simulations are needed to understand these observations.  

	Traditionally, the term wind-fed X-ray binaries has implied Bondi-Hoyle-Lyttleton accretion while disk-fed systems imply Roche-lobe overflow.  Observations of accreting X-ray binaries are beginning to depict a much more complicated picture with systems that  include disk wind accretion and direct wind accretion systems.  
\clearpage
\acknowledgments{\bf{Acknowledgements}}

We acknowledge Slawomir Suchy for the use of his results from his observations of OAO 1657-415 by {\it RXTE}.  M. F. and A. C. acknowledge support from NASA grants NNX08AW06G and NNX11AE24G.  P. Jenke was supported by an appointment to the NASA Postdoctoral Program at the Marshall Space Flight Center, administered by Oak Ridge Associated Universities through a contract with NASA.  This work was supported by the \emph{Fermi} Guest Investigator program and by the Astrophysics Data Analysis Program (ADAP.)
\bibliographystyle{astron}

\bibliography{oao_1657_120827}
\IfFileExists{\jobname.bbl}{}
 {\typeout{}
  \typeout{******************************************}
  \typeout{** Please run "bibtex \jobname" to optain}
  \typeout{** the bibliography and then re-run LaTeX}
  \typeout{** twice to fix the references!}
  \typeout{******************************************}
  \typeout{}
 }

\end{document}